\documentclass[10pt,dvips,a4paper]{article}
\usepackage{epsfig}

\usepackage{amssymb}
\usepackage{amsmath}
\usepackage{setspace}

\begin{document}

\renewcommand{\title}{A parametric study of the meso-scale modelling of concrete subjected to cyclic compression}

\begin{center} \begin{LARGE} \textbf{\title} \end{LARGE} \end{center}

\begin{center}
Rasmus Rempling$^1$ and Peter Grassl$^{2*}$\\

$^1$~Chalmers University of Technology, G\"{o}teborg, Sweden\\
$^2$~University of Glasgow, UK \vspace{1cm}\\

$^*$ Corresponding author.\\
Address: Department of Civil Engineering, University of Glasgow, United Kingdom\\
Email: grassl@civil.gla.ac.uk\\
Phone: +44 141 330 5208\\
Fax: +44 141 330 4557\vspace{1cm}\\
\end{center}

Preprint of Computers and Concrete, Vol. 5, Issue 4, pp. 359-373, 2008


\section*{Abstract}
The present parametric study deals with the meso-scale modelling of concrete subjected to cyclic compression, which exhibits hysteresis loops during unloading and reloading.
Concrete is idealised as a two-dimensional three-phase composite made of aggregates, mortar and interfacial transition zones (ITZs).
The meso-scale modelling approach relies on the hypothesis that the hysteresis loops are caused by localised permanent displacements, which result in nonlinear fracture processes during unloading and reloading.
A parametric study is carried out to investigate how aggregate density and size, amount of permanent displacements in the ITZ and the mortar, and the ITZ strength influence the hysteresis loops obtained with the meso-scale modelling approach.

\section{Introduction}
Concrete subjected to cyclic compressive loading is characterised by a strongly nonlinear response.
Unloading and reloading is accompanied by nonlinear fracture processes on the meso-scale, such as debonding and slip between inclusion and matrix, which result in hysteresis loops \cite{SinGerTul64, KarJir69, Mie84, CorHorRei86}.
Furthermore, repeated loading of concrete subjected to cyclic loading may lead to failure although the stress applied is below the strength obtained for monotonic loading.

Many phenomenological constitutive models were developed over the last 30 years, which can describe the hysteresis loops and the fatigue strength observed in experiments. Examples are constitutive models based on bounding surface models \cite{YanDafHer85, VoyAbu94, PanKryOrt99} or special combinations of damage and plasticity \cite{RagNorMaz00}.
These phenomenological constitutive models have only limited predictive capabilities since they are based on curve fitting and not on the mechanics of the underlying physical processes. Therefore, they cannot predict reliably the response of concrete outside the range of experimental results available. 

Recently, the present authors have proposed a damage-plasticity interface model to describe the response of concrete subjected to cyclic loading \cite{GraRem08}. In this approach concrete was idealised as a three phase-composite consisting of aggregates, mortar and interfacial transition zones (ITZs). A special lattice model was used to descretise the different phases \cite{BolSai98, BolSuk05}.
The meso-scale modelling of hysteresis loops and fatigue strength was based on the assumption that the nonlinear fracture processes during unloading and reloading are caused by localised permanent displacements. The amount of permanent displacements was controlled by a damage-plasticity interface model which was applied to the mortar and ITZ phase. Aggregates were assumed to respond elastically.
This meso-scale modelling approach may have better predictive qualities than macroscopic phenomenological models, since the macroscopic response can be designed by changing the response of the meso-scale constituents.
Nevertheless, each of the three phases has a complex heterogeneous microstructure, which again requires numerical modelling by means of phenomenological models or multiscale analysis in which the mechanical response of the three phases is determined by their micro-structures \cite{BudOco76, HeuLemUlm05, DorKonUlm06}.

In the present work, this meso-scale modelling direction is further pursued and a parametric study is performed to investigate the influence of several model parameters on the description of hysteresis loops. The first two parameters are the volume fraction of aggregates and the aggregate size, which are related to the random discretisation of the composite. Furthermore, three material parameters, which influence the amount of permanent displacements in the mortar and the ITZ and control the ratio of the strength of mortar and interfacial transition zone, are investigated. This parametric study is similar to other parametric studies presented in the literature on the modelling of concrete \cite{PraMie03, LilMie03}.
Nevertheless, the present work is focused on the meso-scale modelling of concrete subjected to cyclic loading.
This is a direction which has not been pursued before to the authors knowledge.

\section{Modelling approach} \label{sec:model}
In the present study the response of concrete subjected to cyclic loading is analysed with a damage-plasticity interface model, which was developed in \cite{GraRem08}. In the following paragraphs, the modelling approach is briefly reviewed to introduce the parameters, which are investigated in the present study.
The interface model relies on a combination of a plasticity model formulated in the effective stress space and isotropic damage mechanics.
For the two-dimensional version of the model, the displacement jump $\mathbf{u}_{\rm c} = \left(u_{\rm n}, u_{\rm s}\right)^T$ at an interface is considered, which is transferred into strains $\boldsymbol{\varepsilon} = \left(\varepsilon_{\rm n}, \varepsilon_{\rm s}\right)^T$ by means of the length $h$ as

\begin{equation}
\boldsymbol{\varepsilon} = \dfrac{\mathbf{u}_{\rm c}}{h}
\end{equation}

The strains are related to the nominal stress $\boldsymbol{\sigma} = \left(\sigma_{\rm n}, \sigma_{\rm s}\right)^T$ as 
\begin{equation} \label{eq:totStressStrain}
\boldsymbol{\sigma} = \left(1-\omega \right) \mathbf{D}_{\rm e} \left(\boldsymbol{\varepsilon} - \boldsymbol{\varepsilon}_{\rm p}\right) = \left(1-\omega\right) \bar{\boldsymbol{\sigma}}
\end{equation}
$\omega$ is the damage variable, $\mathbf{D}_{\rm e}$ is the elastic stiffness,  $\boldsymbol{\varepsilon}_{\rm p} = \left(\varepsilon_{\rm pn}, \varepsilon_{\rm ps}\right)^T$ is the plastic strain and $\bar{\boldsymbol{\sigma}} = \left(\bar{\sigma}_{\rm n}, \bar{\sigma}_{\rm s}\right)^T$ is the effective stress.
The elastic stiffness is 
\begin{equation}
\mathbf{D}_{\rm e} = \begin{Bmatrix} E & 0\\
  0 & \gamma E
\end{Bmatrix}
\end{equation}
where $E$ and $\gamma$ are model parameters controlling both the Young's modulus and Poisson's ratio of the material.

The small strain plasticity model is based on the effective stress $\bar{\boldsymbol{\sigma}}$ and consists of the yield surface, flow rule, evolution law for the hardening parameter and loading and unloading conditions.
A detailed description of the components of the plasticity model is presented in \cite{GraRem08}.
The initial yield surface is determined by the tensile strength $f_{\rm t}$, by the ratio $s$ of the shear and tensile strength, and the ratio $c$ of the compressive and tensile strength.
The evolution of the yield surface during hardening is controlled by the model parameter $\mu$, which is defined as the ratio of permanent and reversible inelastic displacements.
The scalar damage part is chosen so that linear stress inelastic displacement laws for pure tension and compression are obtained, which are characterised by the fracture energies $G_{\rm ft}$ and $G_{\rm fc}$. 

The eight model parameters $E$, $\gamma$, $f_{\rm t}$, $s$, $c$, $G_{\rm ft}$, $G_{\rm fc}$ and $\mu$ can be determined from a tensile, shear and compressive test of the material phase.

\begin{figure} [t]
\begin{center}
\epsfig{file=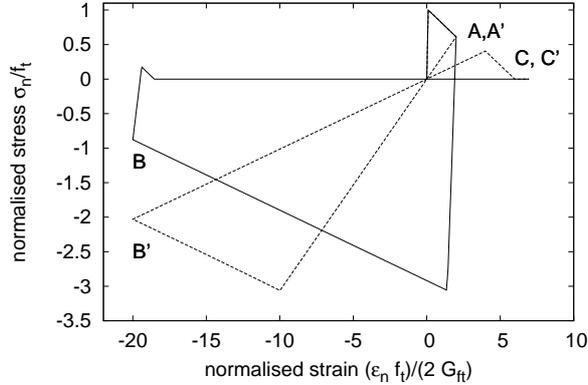, width=8cm}
\end{center}
\caption{Stress-strain response for fluctuating normal strains for  $\mu = 1$ (solid line) and $\mu = 0$ (dashed line).}
\label{fig:constCyclic}
\end{figure}

The constitutive response of the interface model is demonstrated by the stress-strain response for fluctuating normal strains for $\mu = 1$ and $\mu = 0$ (Figure~\ref{fig:constCyclic}).
The normal strain is increased to point $A$ ($A^{'}$). Then the strain is reduced to point $B$ ($B^{'}$) and again increased to point $C$ ($C^{'}$).
The parameter $\mu$ controls the amount of plastic strains.
For $\mu=0$ a pure damage-mechanics response is obtained and the stress-strain curve is unloaded to the origin.
For $\mu = 1$, on the other hand, a pure plasticity model is obtained.
The unloading is elastic and the compressive strength is reached sooner than for $\mu = 0$.
However, the magnitude of the compressive strength is similar for $\mu = 1$ and $\mu=0$, since the plasticity model is based on the effective stress.
The constitutive model describes the loss of stiffness and permanent strains, but it does not represent the hysteresis loops for unloading and reloading.
Instead, these loops are obtained from the structural analysis, which is discussed in the following section.

\begin{figure} [b!]
\begin{center}
\begin{tabular}{c}
\epsfig{file = 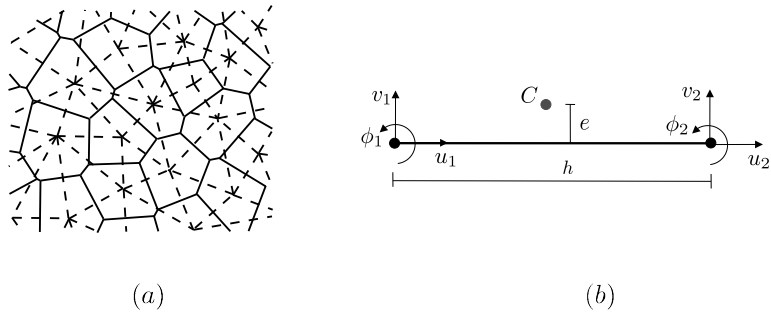, width =10.cm} 
\end{tabular}
\end{center}
\caption{Discretisation: (a) Lattice elements (dashed lines) connecting centroids obtained with the Voronoi tesselation (solid lines). (b) Degrees of freedom $u_1$, $v_1$, $\phi_1$, $u_2$, $v_2$ and $\phi_2$ of the lattice element of length $h$ in the local coordinate system. The point $C$ at which the interface model is evaluated is in the center of the polygon facet at a distance $e$ from the center of the lattice element.}
\label{fig:vorDel}
\end{figure}

The interface model is applied to the two-dimensional plane stress meso-scale analysis of concrete.
A special lattice-type model developed by \cite{Kaw77, MorSawKob93, BolSai98} is used to discretise the domain.
The  domain is decomposed into polygons by means of the Voronoi tessellation \cite{Aur91} (Figure~\ref{fig:vorDel}a). 
The lattice elements connect the centroids of the polygons.
Each node possesses two translations and one rotation shown in the local coordinate system in Figure~\ref{fig:vorDel}b.
The degrees of freedom $\mathbf{u}_{\rm e} = \left\{u_1, v_1, \phi_1, u_2, v_2, \phi_2\right\}^{\rm T}$ of the two nodes of the lattice element are related to the displacement discontinuities $\mathbf{u}_{\rm c} = \left\{u_{\rm c}, v_{\rm c}\right\}^{\rm T}$ at the mid-point $C$ of the interface by
\begin{equation}
\mathbf{u}_{\rm c} = \mathbf{B} \mathbf{u}_{\rm e}
\end{equation}
where
\begin{equation}\label{eq:bMatrix}
\mathbf{B} = \begin{bmatrix}
-1 & 0 & e & 1 & 0 & -e\\
0 & -1 & -h/2 & 0 & 1 & -h/2
\end{bmatrix}  
\end{equation}

In Eq.~(\ref{eq:bMatrix}), the variable $h$ is the element length and $e$ is the eccentricity, defined as the distance between the mid-points of the lattice element and the corresponding polygon facet, respectively (Figure~\ref{fig:vorDel}).
The displacements $\mathbf{u}_{\rm c}$ at the point $C$ are transformed into strains 
$\boldsymbol{\varepsilon} = \left\{ \varepsilon_{\rm n}, \varepsilon_{\rm s} \right\}^{T} = \mathbf{u}_{\rm c} /h$ (Figure~\ref{fig:vorDel}b).
The strain $\boldsymbol{\varepsilon}$ is related to the stress $\boldsymbol{\sigma} = \left\{\sigma_{\rm n}, \sigma_{\rm s} \right\}^{T}$ by means of the two-dimensional version ($\sigma_{\rm t} = 0$) of the damage-plasticity interface model presented earlier.
The element stiffness is
\begin{equation}
\mathbf{K}_{\rm e} = \dfrac{A}{h} \mathbf{B}^{\rm T} \mathbf{D} \mathbf{B}
\end{equation}
where $A$ is the length of the facet.

The nodes of the lattice elements are placed sequentially in the domain to be analysed. 
The coordinates of each node are determined randomly and a minimum distance $d_{\rm m} = 3$~mm is enforced iteratively between the nodes \cite{ZubBaz87}.
For this iterative process, the number of vertices $n$ for a chosen domain $A_{\rm d}$ and the minimum distance $d_{\rm m}$ determine the distribution of lengths of lattice elements. This relationship can be expressed in the form of a density
\begin{equation}
\rho = \dfrac{n d_{\rm m}^2}{A_{\rm d}}
\end{equation}
The resulting irregular arrangement of lattice elements is more suitable for fracture analysis, since the fracture patterns obtained are less sensitive to the arrangement of the lattice elements.

The meso-structure of the heterogeneous material is discretised by lattice elements perpendicular to the boundary between the cylindrical inclusions and the matrix. 
The diameters of the inclusions are determined randomly from the cumulative distribution function
\begin{equation}\label{eq:cdfFuller}
P_{\rm f} = \dfrac{1-\left(\dfrac{\phi_{\rm min}}{d}\right)^{2.5}}{1-\alpha^{-2.5}}
\end{equation}
where $P_{\rm f}$ is the probability of the occurrence of the inclusion diameter $d$. Furthermore, $\alpha = \phi_{\rm max}/\phi_{\rm min}$, where $\phi_{\rm min}$ and $\phi_{\rm max}$ are the minimum and maximum diameters, respectively \cite{CarCorPuz04}. 
A pseudo-random number generator is used to generate probabilities from which the diameter $d$ is determined by means of the inverse of Eq.~(\ref{eq:cdfFuller}).
This procedure is repeated until the chosen volume fraction $\rho_{\rm a}$ of circular inclusions is reached.
The inclusions are placed sequentially by means of randomly generated coordinates within the area of the specimen. For each set of generated coordinates, it is checked that no overlap with previously placed inclusions occurs. However, overlap with boundaries is permitted, i.e. no boundary effect is considered.
 
\section{Parametric study}
The present parametric study deals with the modelling of concrete subjected to cyclic loading.
A three-dimensional concrete cylinder subjected to cyclic uniaxial compression tested by Sinha, Gerstle and Tulin \cite{SinGerTul64} was idealised by means of a two-dimensional meso-scale plane stress model.
Concrete was represented by a three-phase composite consisting of mortar, aggregates and interfacial transition zones between the two phases.

\begin{figure}
\begin{center}
\begin{tabular}{cccc}
\epsfig{file=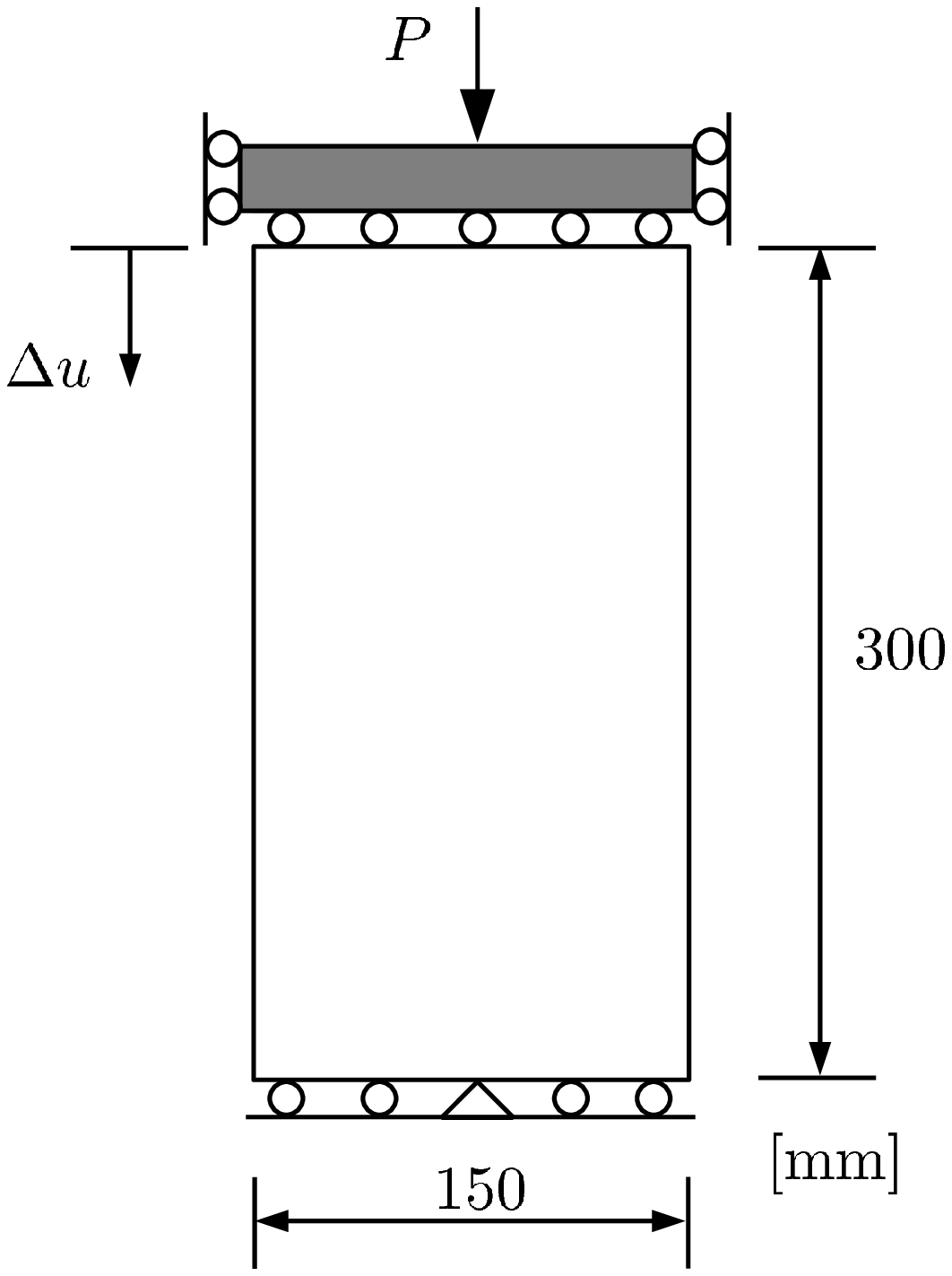, width= 5cm} & \epsfig{file=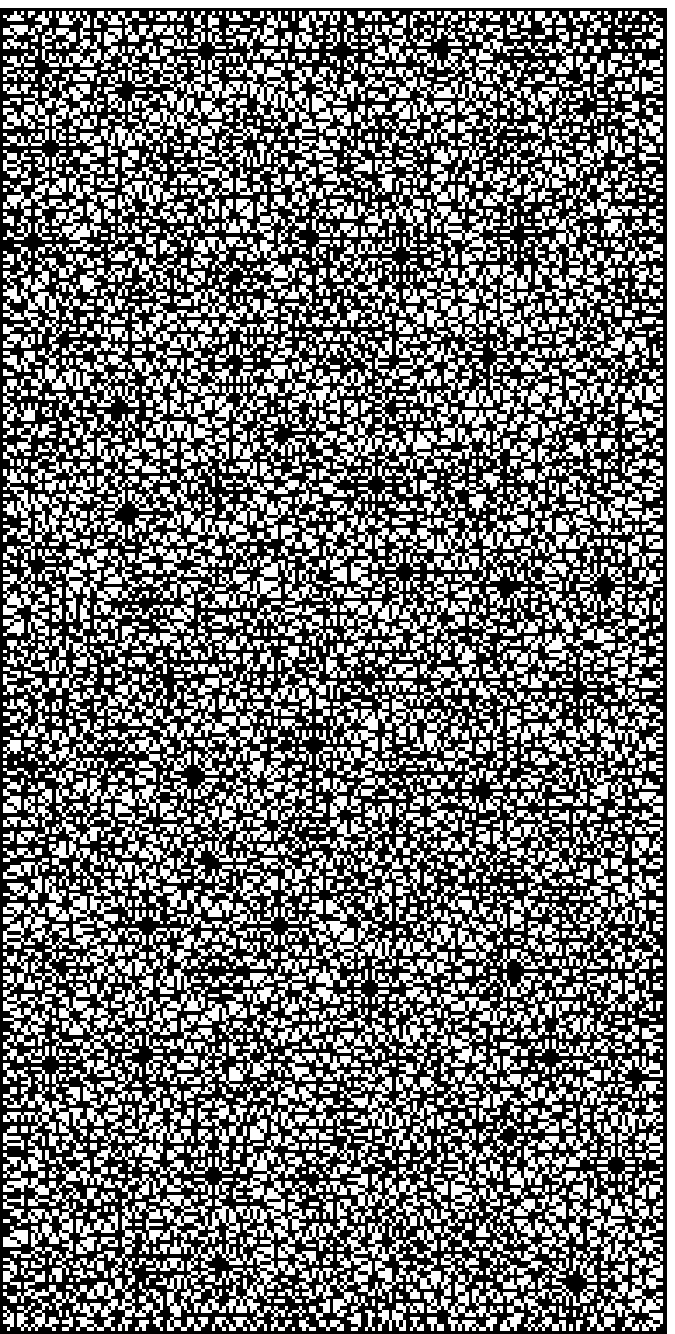, width= 3cm} & \epsfig{file=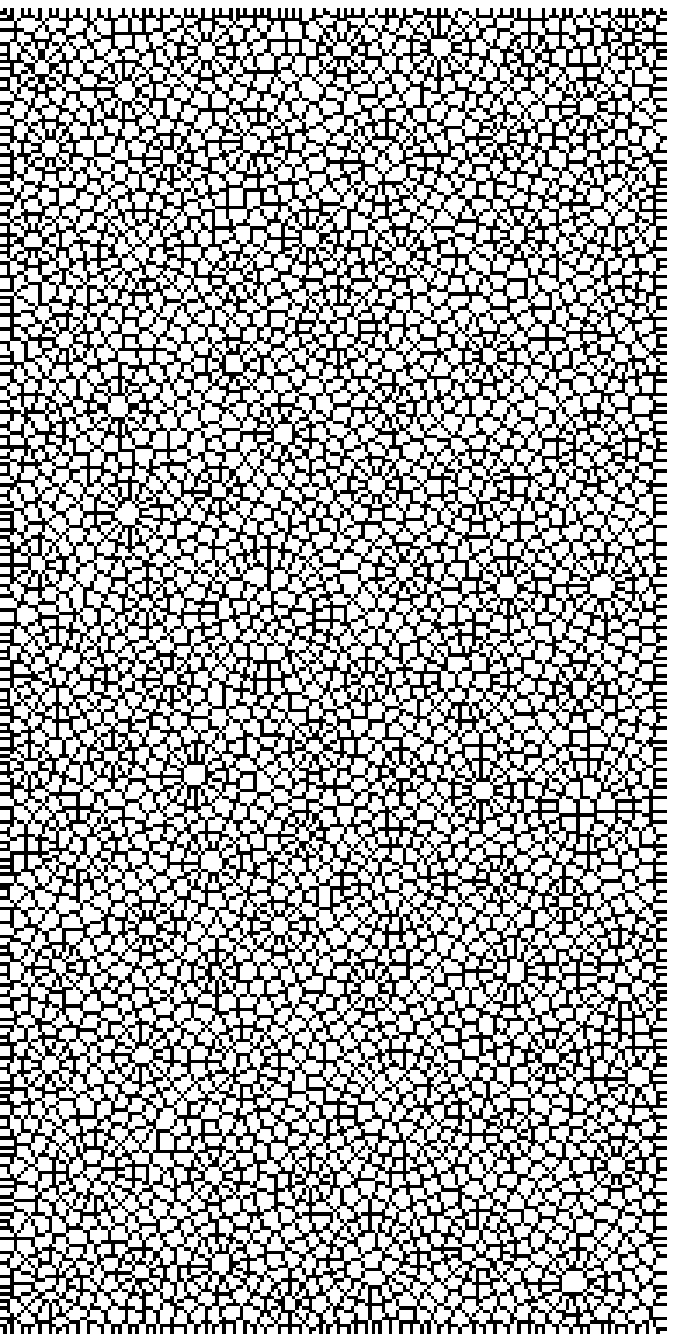, width= 3cm} & \epsfig{file=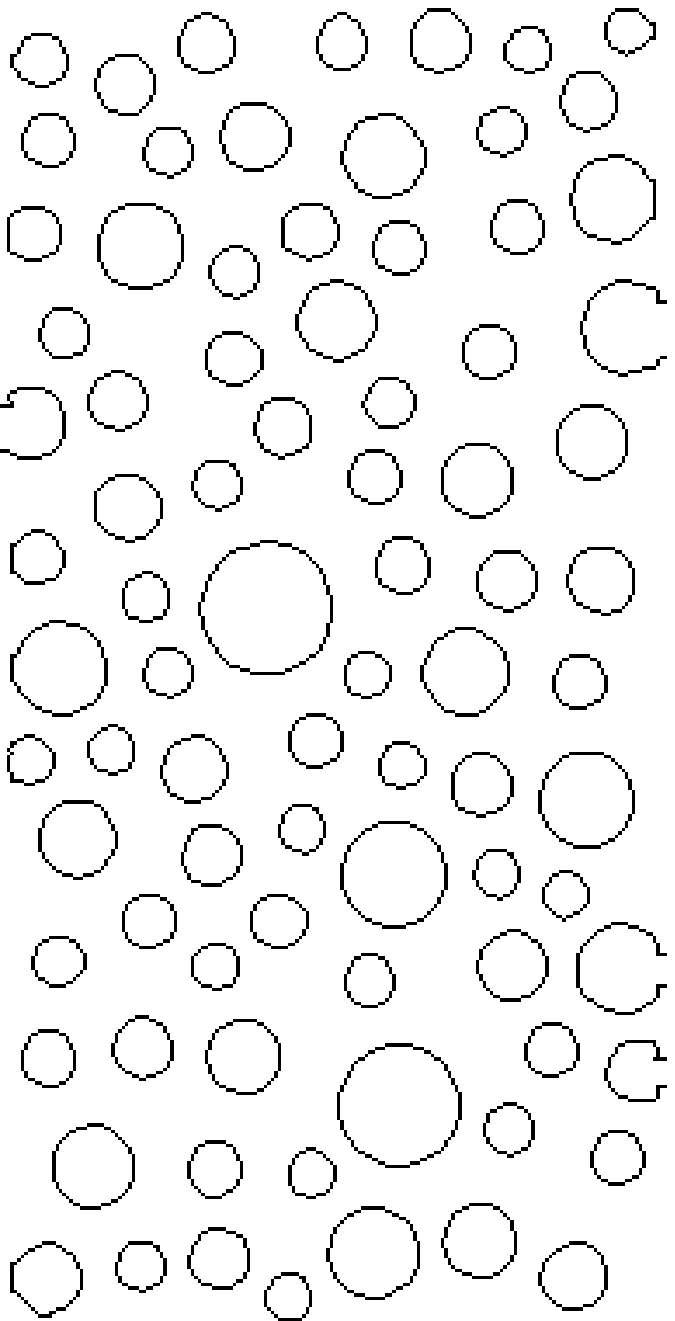, width= 3cm}\\
(a) & (b) & (c) & (d) 
\end{tabular}
\end{center}
\caption{Concrete in uniaxial compression: a) Geometry and loading setup, (b) Lattice (c) Voronoi polygons (d) Boundaries of the inclusions.}
\label{fig:geom}
\end{figure}

The meso-structure was discretised by means of a random lattice based on the Voronoi tesselation of the domain.
The mesh was generated with a vertex density of $\rho = 0.6$ and a minimum distance of $d_{\rm m}=3$~mm.
This vertex density is close to the saturated vertex density which was determined by Bolander and Saito in \cite{BolSai98} as $\rho = 0.68$.
The aggregate distribution was obtained with $d_{\rm max} = 32$~mm, $d_{\rm min} = 10$~mm and $\rho_{\rm a} = 0.3$.
The geometry, loading setup and one exemplary mesh are shown in Figure~\ref{fig:geom}.

\begin{figure}
\begin{center}
\begin{tabular}{c}
  \epsfig{file=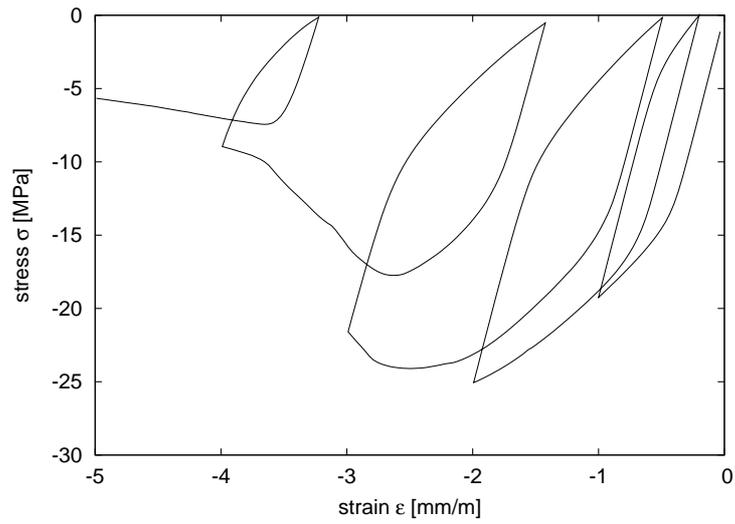, width =10cm} \\
(a)\\
  \epsfig{file=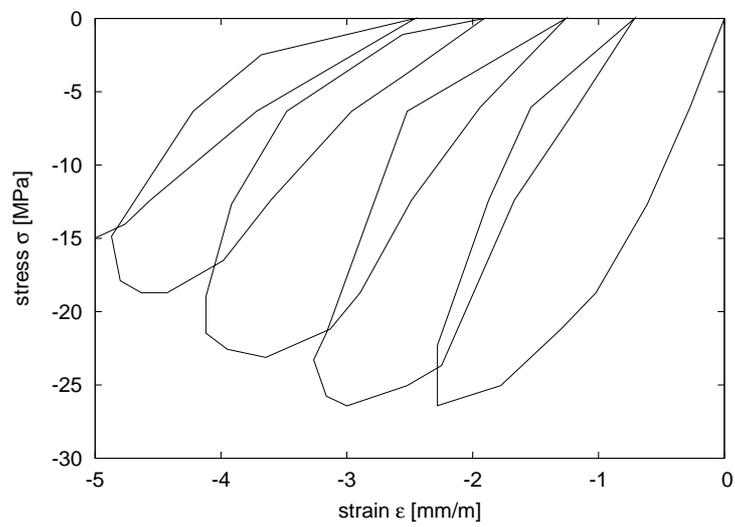, width =10cm}\\
(b)
\end{tabular}
\end{center}
\caption{Stress-strain responses obtained from (a) the meso-scale analysis and from (b) experiments reported in \protect \cite{SinGerTul64}.}
\label{fig:cyclicHpFit}
\end{figure}

\begin{figure}
\begin{center}
  \epsfig{file=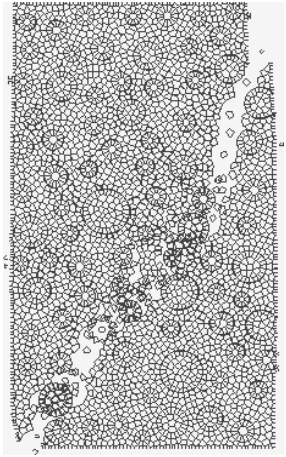, height =6cm} 
\end{center}
\caption{Deformations magnified by a factor of five shown for the monotonic loading.}
\label{fig:monoProcess}
\end{figure}

The concrete specimen subjected to monotonic and cyclic compression was analysed with the model parameters shown in Table~\ref{tab:origParam}.
\begin{table}[t]
\caption{Model parameters.}
\label{tab:origParam}
\vspace{6pt} \center
\begin{tabular}{ccccccccc}
  Phase & $E$ [GPa] & $\gamma$ & $f_{\rm t}$ [MPa] & $s$ & $c$ & $G_{\rm ft}$ [J/mm$^2$] & $G_{\rm fc}$ [J/mm$^2$] & $\mu$ \\\hline
  Mortar & 30 & 0.33 & 5 & 2 & 20 & 400 & 40000 & 1 \\
  Interface & 48 & 0.33 & 1 & 2 & 20 & 80 & 8000 & 1 \\
  Aggregate & 120 & 0.33 & - & - & - & - & - & -
  \\
\end{tabular}
\end{table}
The axial average stress was determined as $\sigma = P/d$, where $P$ is the axial force, $\Delta u$ is the axial displacement and $d$ is the width of the specimen.
Correspondingly, the axial average strain was defined as $\varepsilon = \Delta u/L$, where $L$ is the height of the specimen.
The analysis results in the form of the average stress-strain curves for monotonic and cyclic loading are compared to the experimental results in Figure~\ref{fig:cyclicHpFit}.
A reasonable qualitative agreement between the modelling response and the experimental results is obtained.
The deformation pattern in the post-peak regime are presented in Figure~\ref{fig:monoProcess}.

Five parameters, which are likely to have a strong influence on the modelling of concrete subjected to cyclic loading, are studied:
\begin{enumerate}
\item Volume fraction of aggregates.
\item Aggregate size.
\item Ratio of permanent and total inelastic displacements in the ITZ phase.
\item Ratio of permanent and total inelastic displacements in the mortar phase.
\item Ratio of the mortar and interface strength.
\end{enumerate}

\begin{figure}
\begin{center}
\begin{tabular}{ccccc}
  \epsfig{file=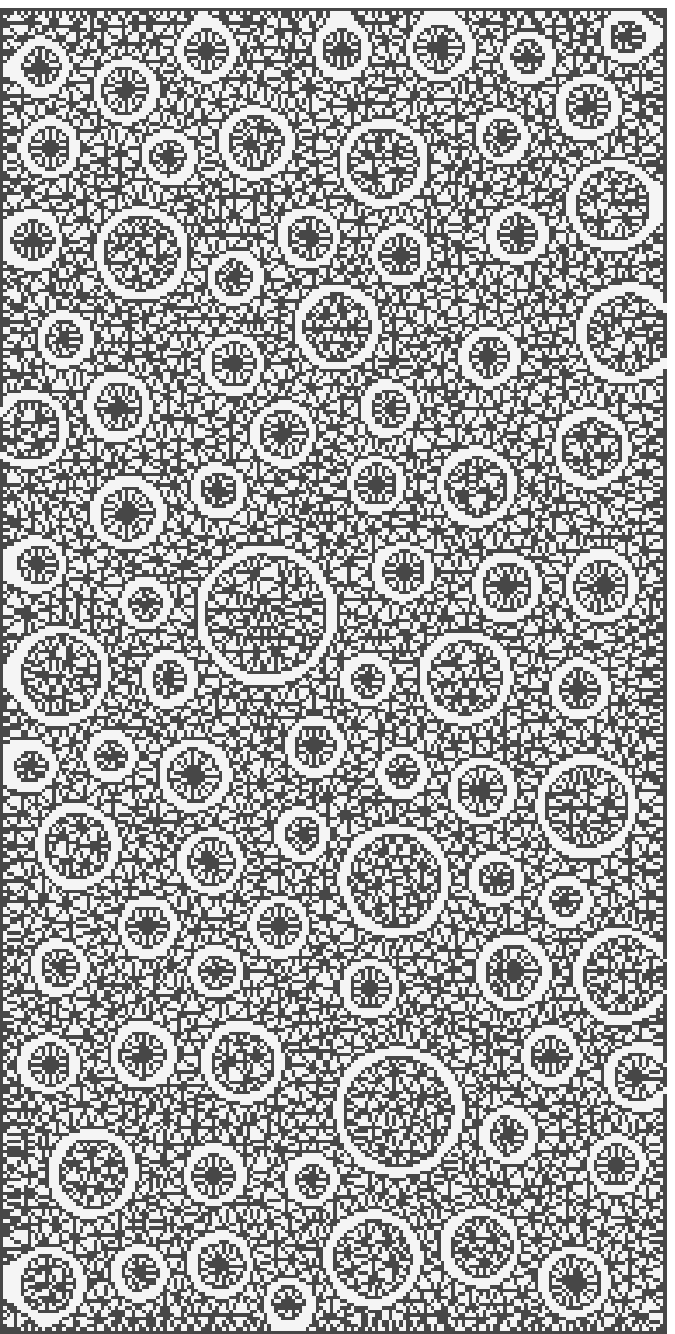, height=6cm} &   \epsfig{file=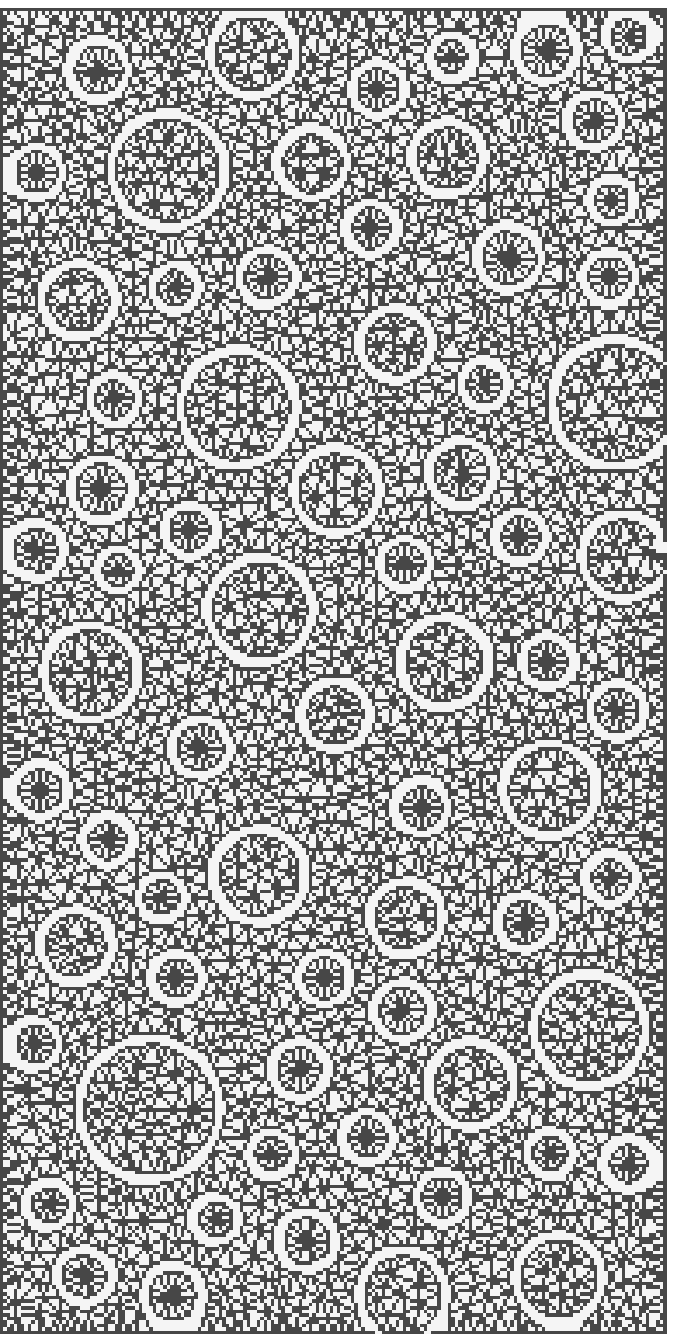, height=6cm} &   \epsfig{file=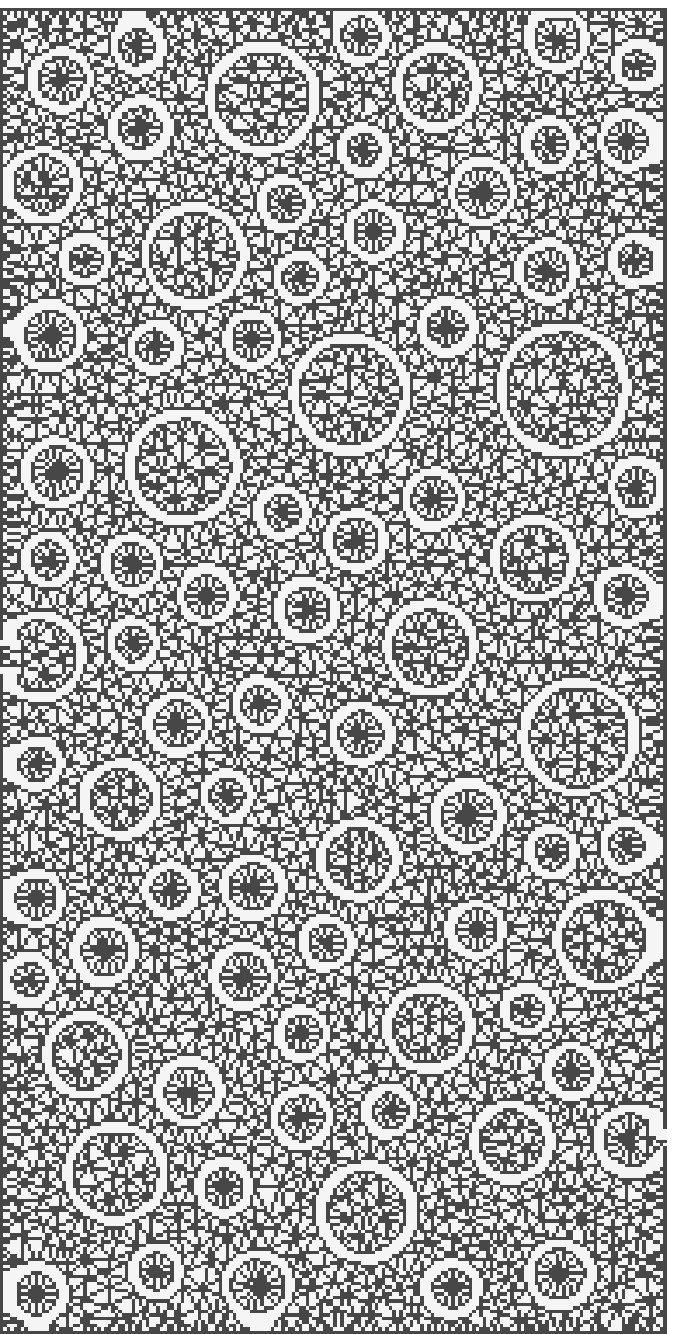, height=6cm} &   \epsfig{file=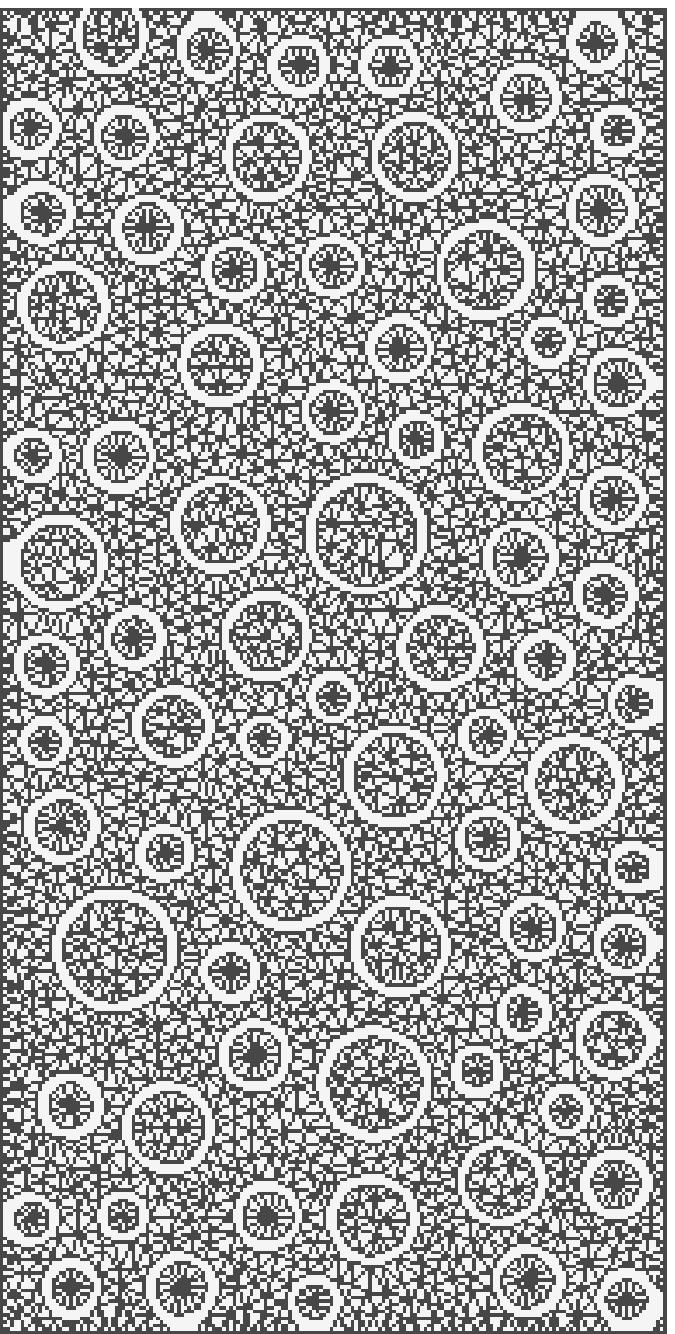, height=6cm} &   \epsfig{file=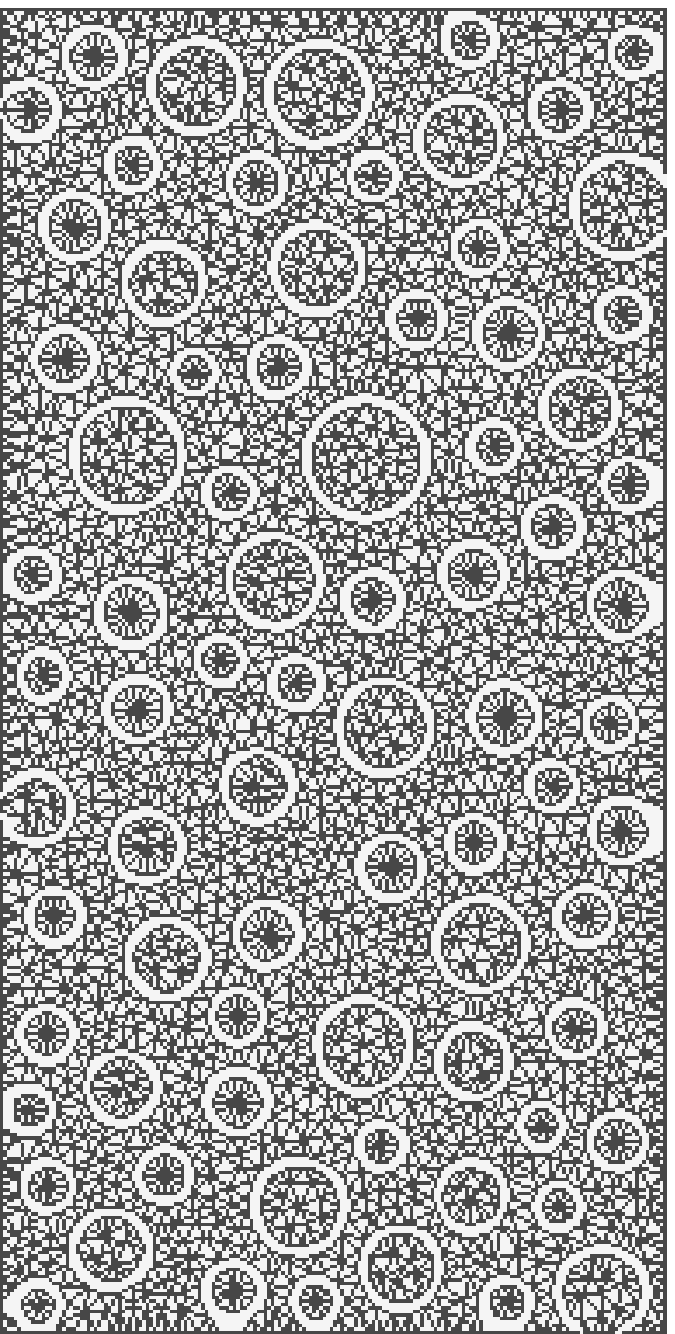, height=6cm}
\end{tabular}
\end{center}
\caption{Five random meshes with the same aggregate density and size distribution. Lattice elements which describe the ITZs between aggregates and mortar are not shown to improve the clarity of the illustration.}
\label{fig:meshCheck}
\end{figure}

The first two parameters, which are related to the aggregate distribution, require random lattice generations.
Random meshes and the random positions of the aggregates influence the average stress-strain response.
Therefore, the influence of different discretisations on the response of monotonic and cyclic compressive loading is initially analysed by means of five analyses with the same aggregate size and distribution, but with random geometry.
The five different meshes are presented in Figure~\ref{fig:meshCheck}.

\begin{figure}
\begin{center}
\epsfig{file=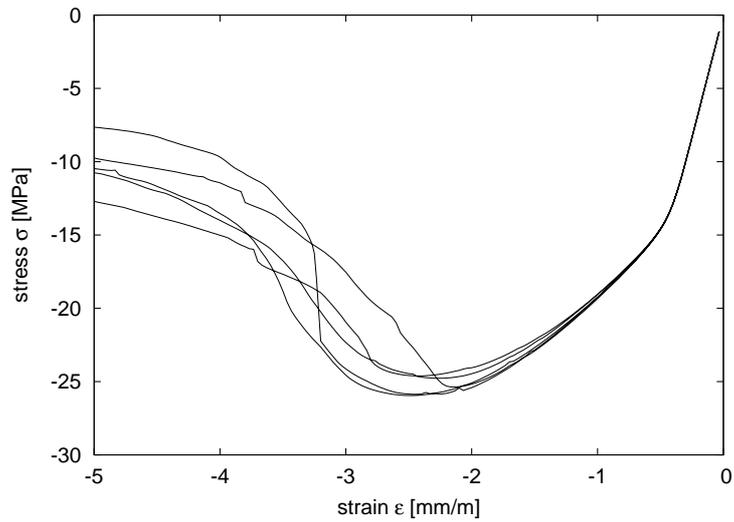, width=10cm}
\end{center}
\caption{Stress-strain responses obtained for monotonic loading for five meshes with different aggregate arrangements and discretisations.}
\label{fig:ldCheckMono}
\end{figure}

\begin{figure}
\begin{center}
\epsfig{file=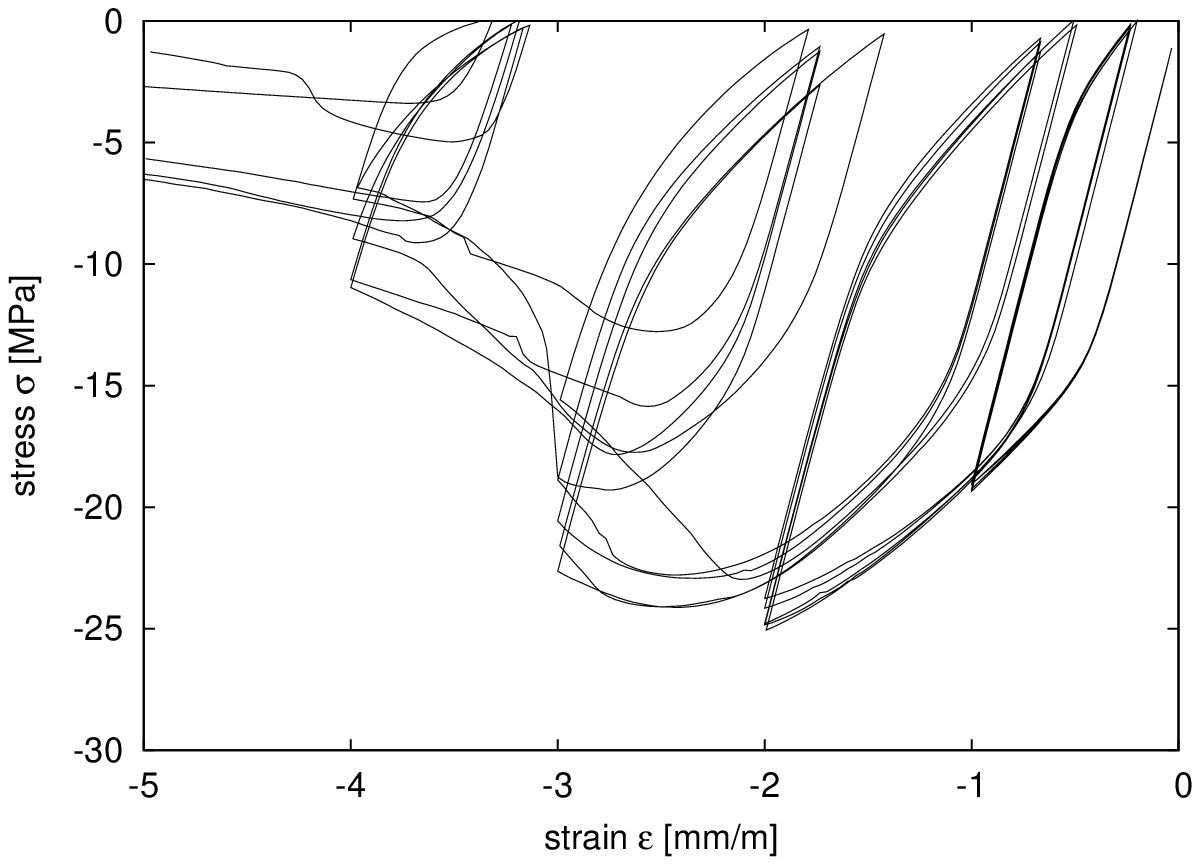, width=10cm}
\end{center}
\caption{Stress-strain responses obtained for cyclic loading for five meshes with different aggregate arrangements and discretisations.}
\label{fig:ldCheckCyclic}
\end{figure}

The response for both monotonic and cyclic compressive loading was analysed.
For cyclic loading, the specimens were subjected to four cycles.
The strain was increased in steps to $\varepsilon_{\rm m} = 0.5 \%$, $1\%$, $1.5\%$, $2\%$, with the strain being reversed after each step so that $\sigma_{\rm m} = 0$.
The same loading history was applied for all the analyses in the present parametric study.
The results for the influence of the random positions of the aggregates and the random orientation of the lattice elements are presented in Figure~\ref{fig:ldCheckMono} for monotonic and in Figure~\ref{fig:ldCheckCyclic} for cyclic loading in the form of average stress-strain relations.
The random discretisation has a small influence on the pre-peak regime, whereby the peak and post-peak regime is strongly influenced.
Nevertheless, the scatter of the simulations is reasonable compared to the one obtained experimentally by van Mier in \cite{Mie84}.

\begin{figure}
\begin{center}
\begin{tabular}{ccc}
\epsfig{file=./Figures/Random/One/mesh.eps, width=3cm} & \epsfig{file=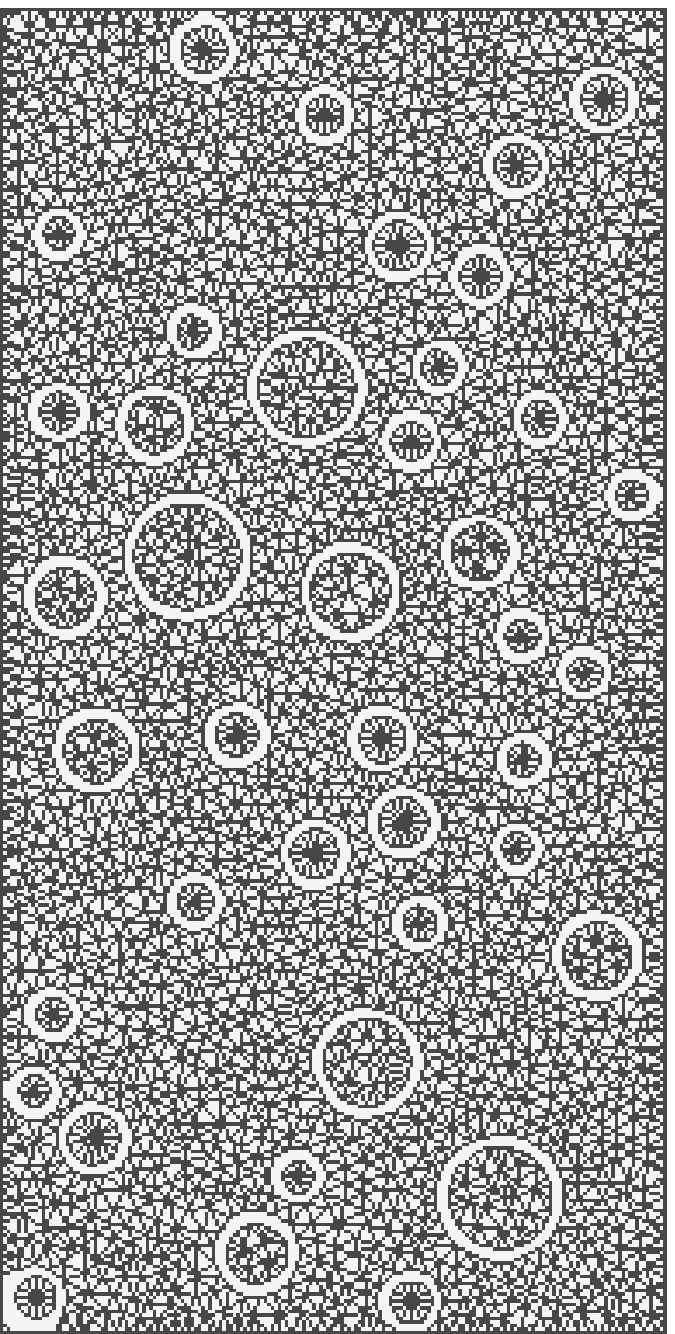, width=3cm} & \epsfig{file=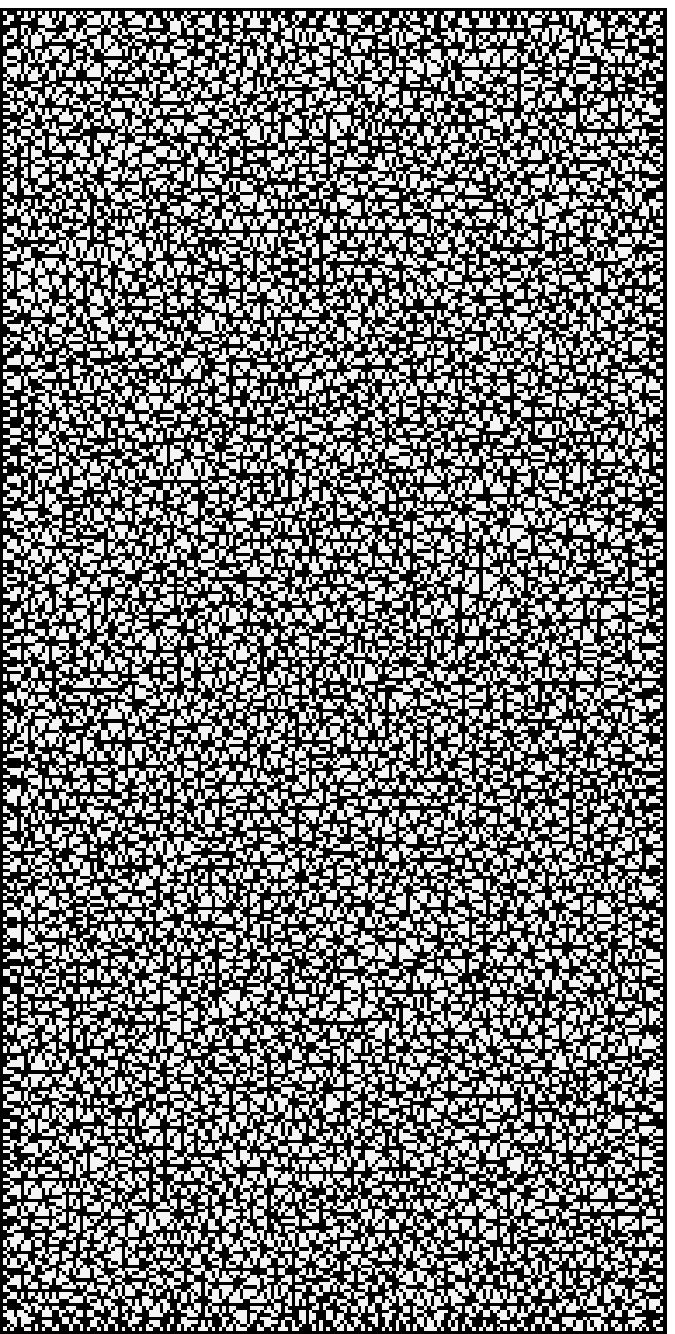,width=3cm} \\
(a) & (b) & (c)
\end{tabular}
\end{center}
\caption{Lattices for aggregate densities of (a) \protect $\rho_{\rm a}=0.3$, (b) \protect $\rho_{\rm a}=0.15$, (c) \protect $\rho_{\rm a}=0$.
Lattice elements which describe the interface between aggregates and mortar are not shown to improve the clarity of the illustration.}
\label{fig:meshpatternds}
\end{figure}

The first of the five parameters is the volume fraction $\rho_{\rm a}$.
The response for three volume fractions $\rho_{\rm a} = 0.3$, $0.15$ and $0$ were analysed.
For $\rho_{\rm a} = 0$ the resulting, stress-strain curves exhibit snap-back in the post-peak regime. 
To avoid numerical problems, the tensile strength of the mortar phase was reduced to $f_{\rm t} = 2.1$~MPa. 
The other model parameters are the same as in Table~\ref{tab:origParam}.
The distribution of the randomly placed aggregates is presented in Figure~\ref{fig:meshpatternds}.

\begin{figure}
\begin{center}
\epsfig{file=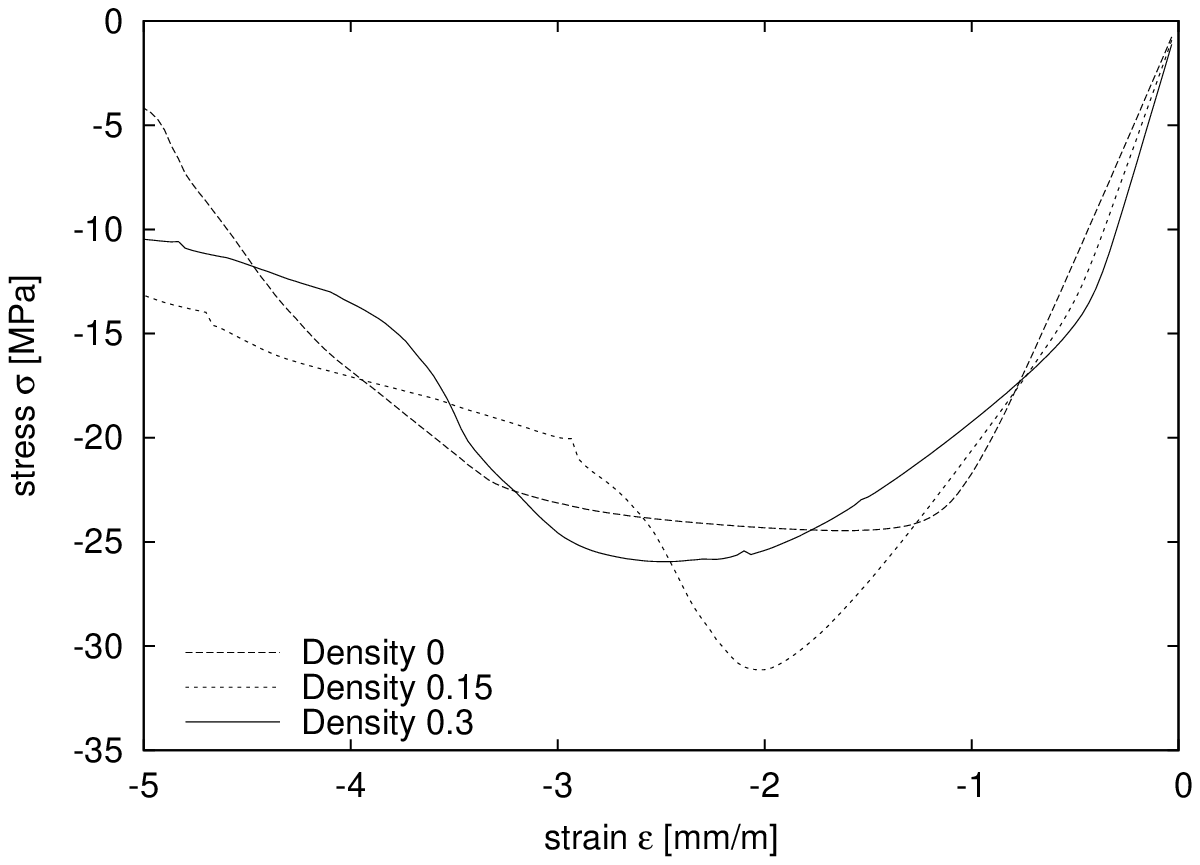,width=10cm}
\end{center}
\caption{Stress-strain responses for monotonic loading for three aggregate densities of $\rho_{\rm a} = 0.3$, $0.15$ and $0$.  The tensile strength of the mortar phase for the specimen with $\rho_{\rm a}=0$ was reduced to $f_{\rm t} = 2.1$~MPa to avoid snap-back in the post-peak regime.}
\label{fig:ldDsMono}
\end{figure}

\begin{figure}
\begin{center}
\epsfig{file=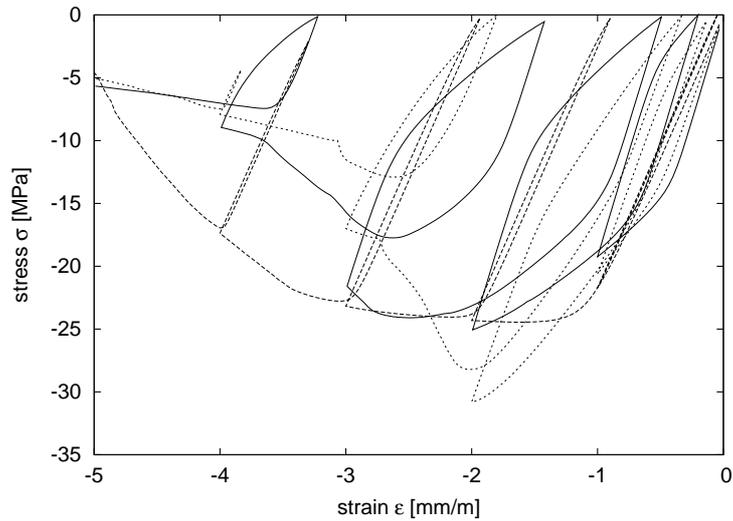, width=10cm}
\end{center}
\caption{Stress-strain responses for cyclic loading for three aggregates densities $\rho_{\rm a} = 0.3$, $0.15$ and $0$. The tensile strength of the mortar phase for the specimen with $\rho_{\rm a}$ was reduced to $f_{\rm t} = 2.1$~MPa to avoid snap-back in the post-peak regime.}
\label{fig:ldDsCyclic}
\end{figure}

The average stress-strain response for the different aggregate volume fractions is presented in Figures~\ref{fig:ldDsMono}~and~\ref{fig:ldDsCyclic} for monotonic and cyclic loading, respectively.
For the monotonic response, a decrease of the aggregate fraction results in an increase of the compressive strength.
This strong influence of the aggregate fraction is explained by the weak interfaces. With an increasing number of interfaces, the strength of the composite is reduced.
For the cyclic response, a decrease of the aggregate volume fraction leads to a decrease of the size of the hysteresis loops.
The occurrence of hysteresis loops is based on the existence of weak zones, in which permanent displacements localise. 
If the number of these localised zones of permanent displacements is reduced, the nonlinear fracture processes during unloading and reloading is decreased and the resulting hysteresis loops are smaller. 

\begin{figure}
\begin{center}
\begin{tabular}{ccc}
\epsfig{file=./Figures/Random/One/mesh.eps,width=3cm} & \epsfig{file=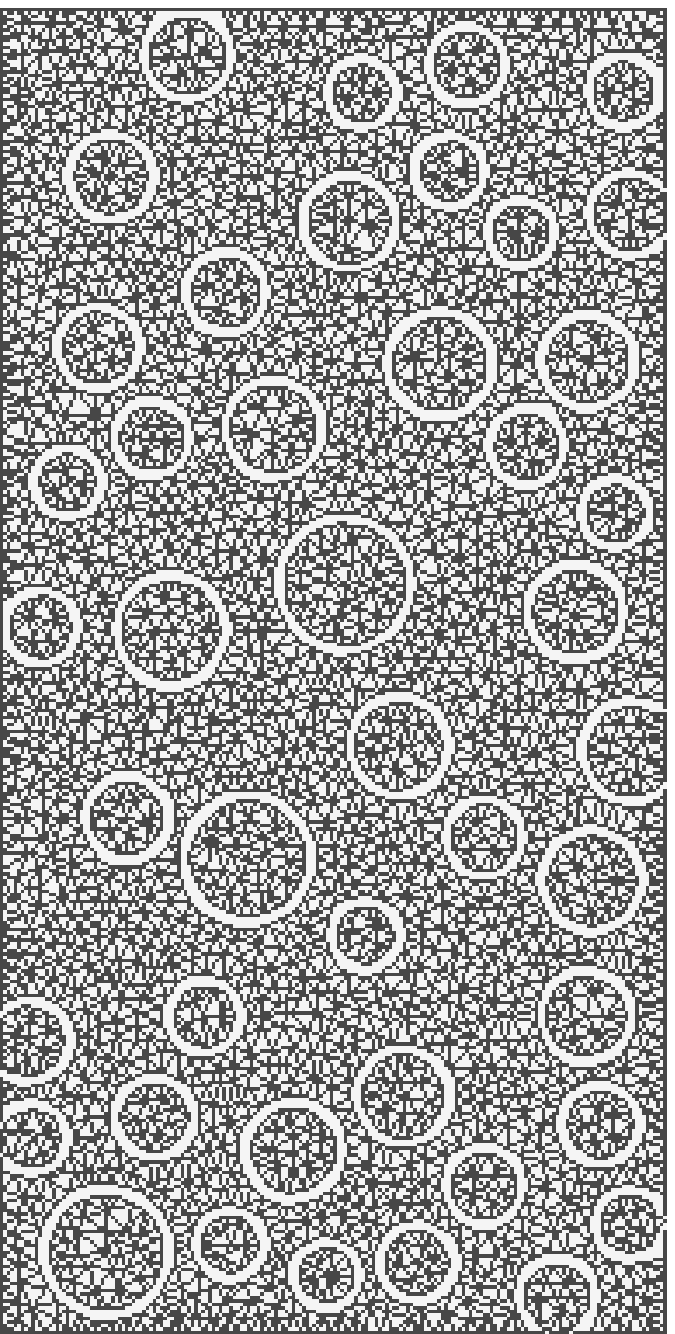,width=3cm} & \epsfig{file=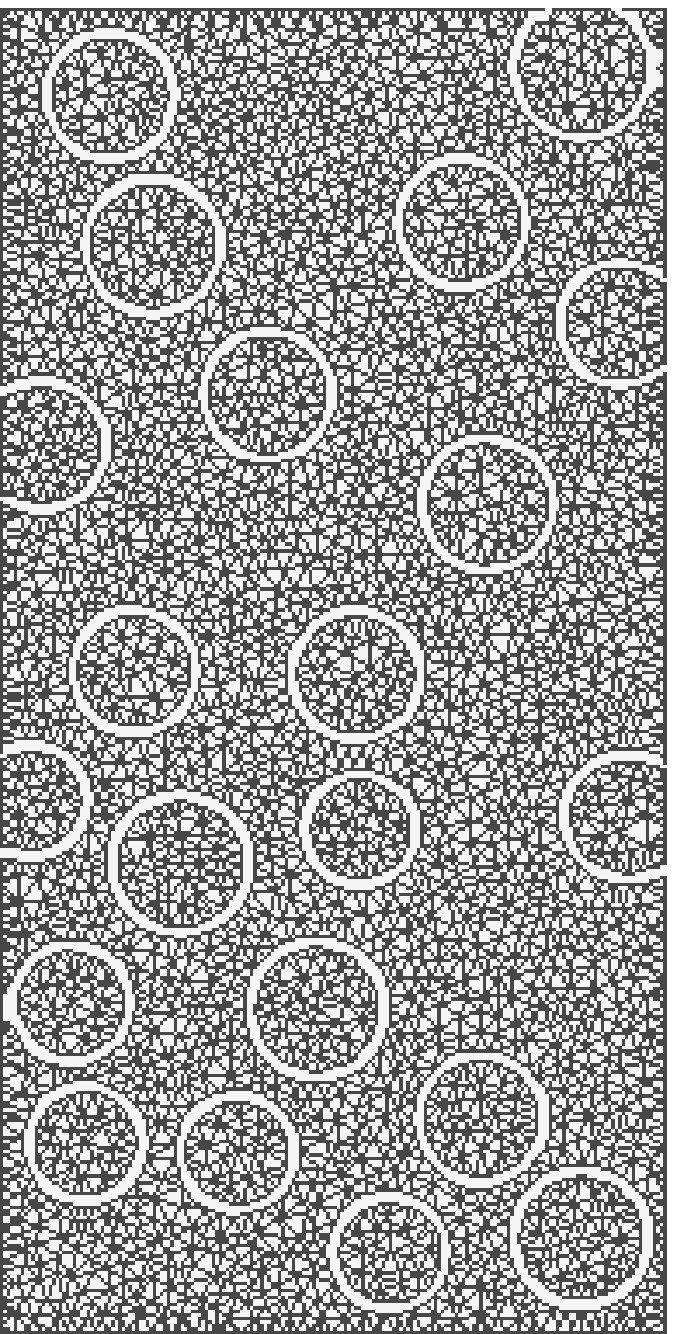,width=3cm}\\
(a) & (b) & (c) 
\end{tabular}
\end{center}
\caption{Lattice elements for the size range of (a) $d_{\rm{max}}/d_{\rm{min}}=3.2$, (b) $d_{\rm{max}}/d_{\rm{min}}=2.1$, (c) $d_{\rm{max}}/d_{\rm{min}}=1.3$.}
\label{fig:meshpatternagg}
\end{figure}

The second parameter studied is the ratio of the maximum and minimum aggregate size $d_{\rm{max}}/d_{\rm{min}}$.
Three size ratios of $d_{\rm{max}}/d_{\rm{min}} = 3.2$, $2.1$ and $1.3$ were chosen. 
For all three cases, the maximum aggregate size was $d_{\rm{max}} = 32$~mm.
The meshes are displayed in Figure~\ref{fig:meshpatternagg}.

\begin{figure}
\begin{center}
\epsfig{file=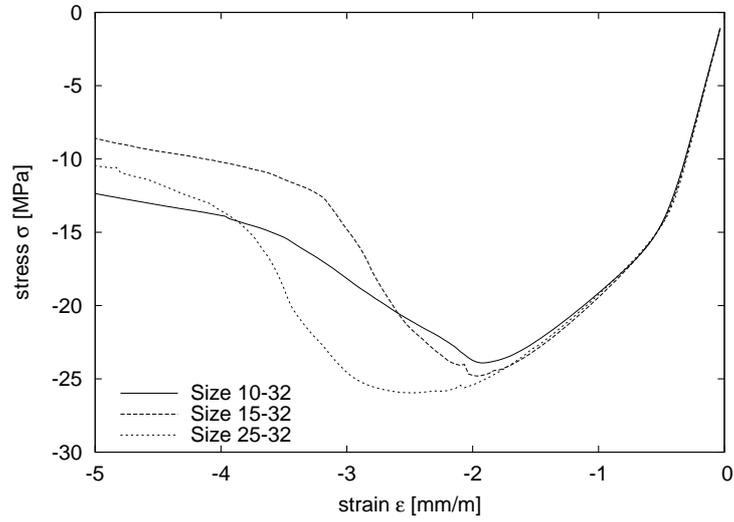,width=10cm}
\end{center}
\caption{Stress-strain responses for monotonic loading for three aggregate-size ranges of $d_{\rm{max}}/d_{\rm{min}} = 3.2$, $d_{\rm{max}}/d_{\rm{min}} = 2.1$ and $d_{\rm{max}}/d_{\rm{min}} = 1.3$.}
\label{fig:ldSizeMono}
\end{figure}

\begin{figure}
\begin{center}
\epsfig{file=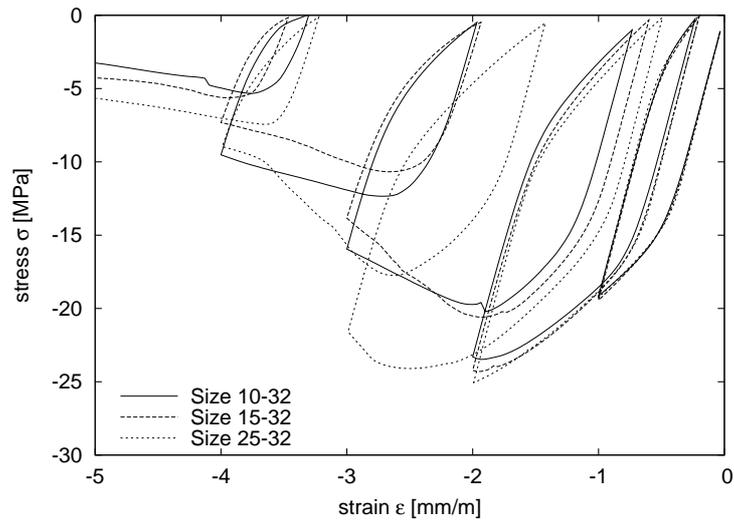, width=10cm}
\end{center}
\caption{Stress-strain responses for cyclic loading for three aggregate-size ranges of $d_{\rm{max}}/d_{\rm{min}} = 3.2$, $d_{\rm{max}}/d_{\rm{min}} = 2.1$ and $d_{\rm{max}}/d_{\rm{min}} = 1.3$.}
\label{fig:ldSizeCycle}
\end{figure}

The average stress-strain curves for  monotonic and cyclic compressive loading for the different aggregate size ratios is shown in Figures~\ref{fig:ldSizeMono} and \ref{fig:ldSizeCycle}. 
The difference between the curves is small and falls in the range of scatters obtained by the initial study of the influence of the discretisation. Thus, the size ratio does not influence the monotonic and cyclic response strongly.
The remaining three parameters are related to the material response and are independent of the discretisation used.
Therefore, the results for these parameters are obtained with the same discretisation.

\begin{figure}
  \begin{center}
    \epsfig{file=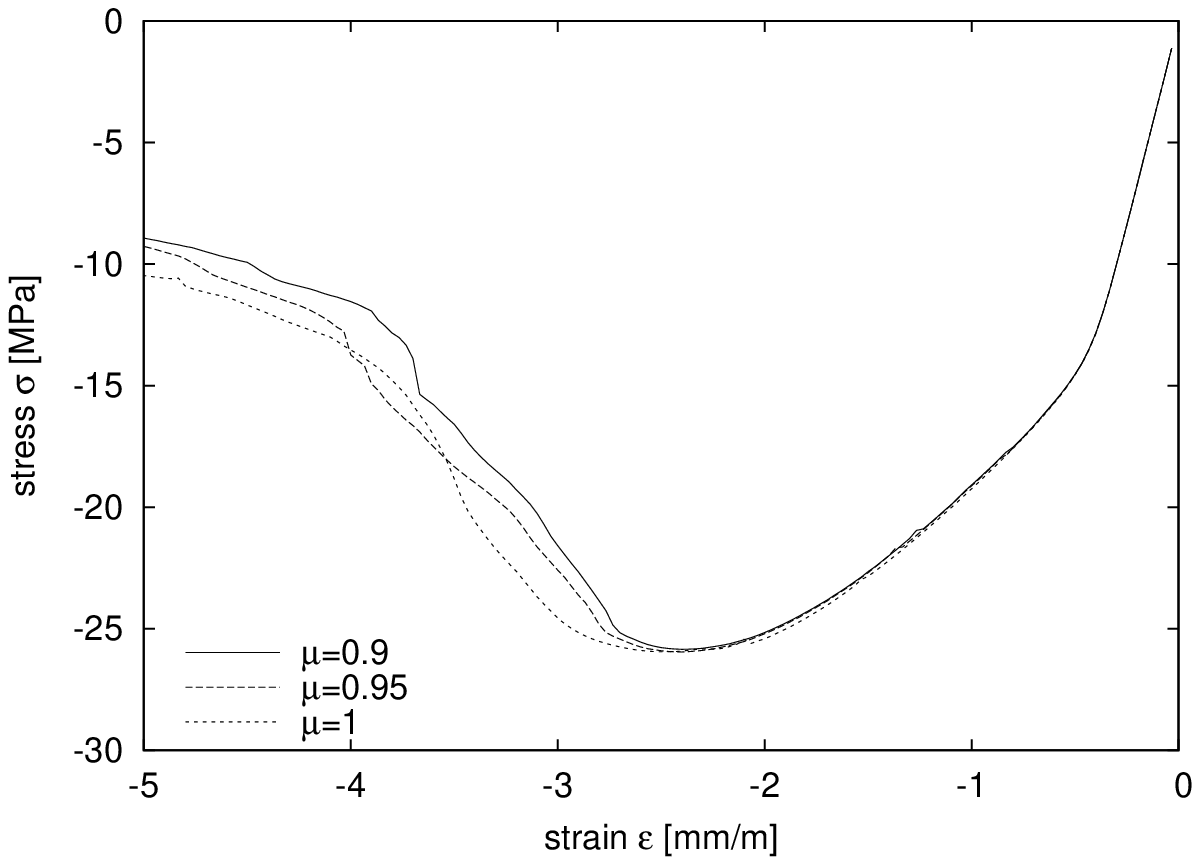,width=10cm}
  \end{center}
  \caption{Stress-strain responses for monotonic loading for $\mu = 1$, $0.95$ and $0.9$ of the ITZ phase.}
\label{fig:ldHpITZMono}
\end{figure}

\begin{figure}
  \begin{center}
    \epsfig{file=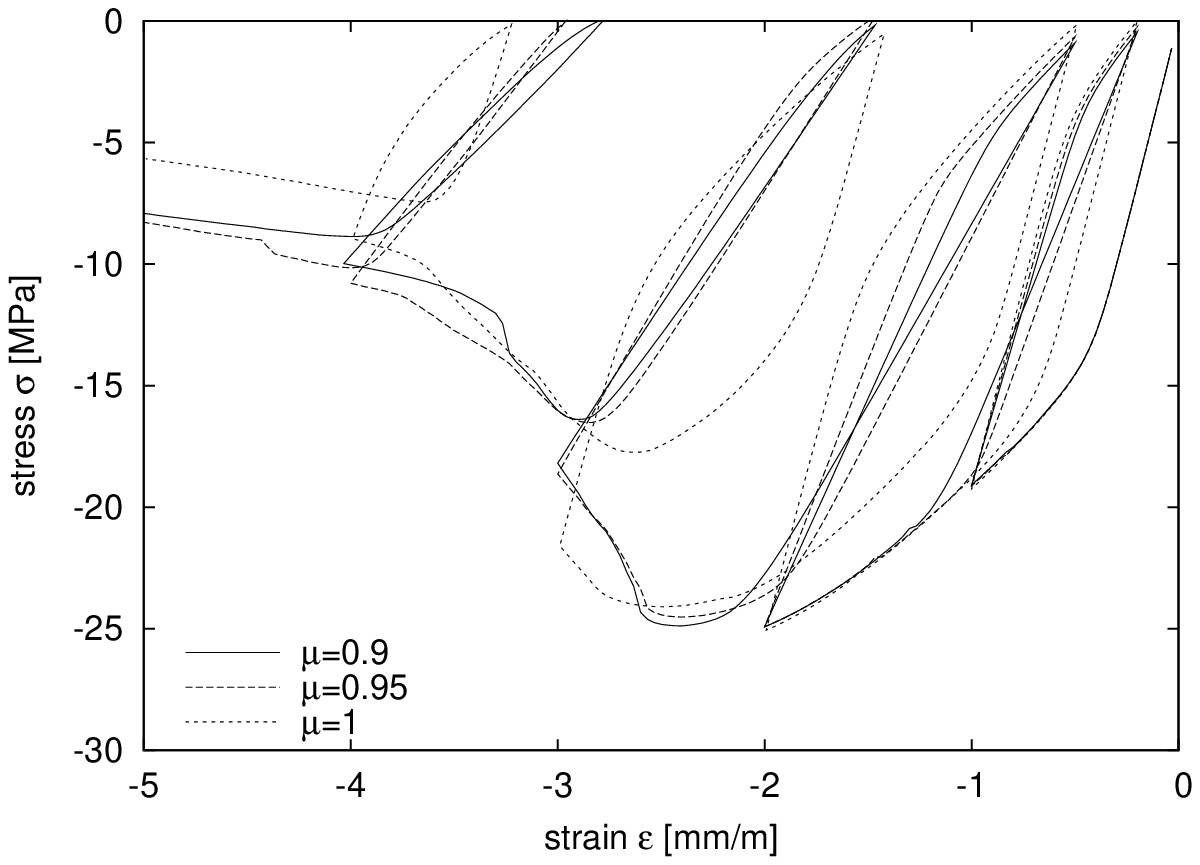,width=10cm}
  \end{center}
  \caption{Stress-strain responses for cyclic loading for $\mu = 1$, $0.95$ and $0.9$ of the ITZ phase.}
  \label{fig:ldHpITZCycl}
\end{figure}

The next two parameters studied are related to the ratio of permanent and total inelastic displacements in the ITZ and mortar phase. 
This ratio is equal to the model parameter $\mu$. 
The present meso-scale approach for cyclic loading is based on the hypothesis that the occurrence of localised permanent displacements is the main cause for the presence of the hysteresis loops. 
Consequently, a reduction of the permanent displacements (decrease of $\mu$) should result in smaller hysteresis loops. 
The ratios $\mu= 1$, $0.95$ and $0.9$ were used.
The monotonic and cyclic stress-strain curves for the three ratios of the ITZ phase are shown in Figure~\ref{fig:ldHpITZMono} and Figure~\ref{fig:ldHpITZCycl}, respectively.
The size of the hysteresis loops decreases with increasing ratio of permanent and total inelastic displacements.
This corresponds to the hypothesis that the hysteresis loops are caused by localised permanent displacement.

\begin{figure}
  \begin{center}
    \epsfig{file=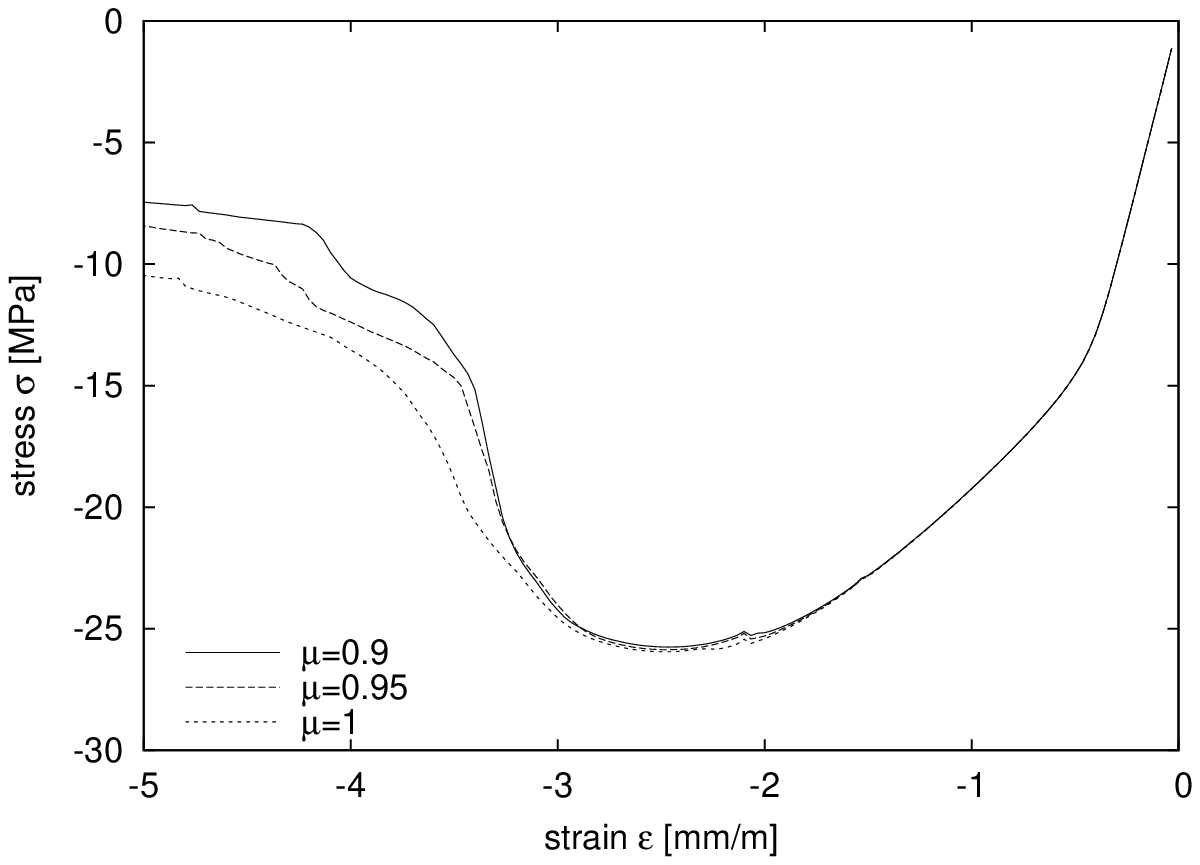,width=10cm}
  \end{center}
  \caption{Stress-strain responses for monotonic loading for $\mu=1$, $0.95$ and $0.9$ of the mortar phase.}
  \label{fig:ldHpMortarMono}
\end{figure}

\begin{figure}
  \begin{center}
    \epsfig{file=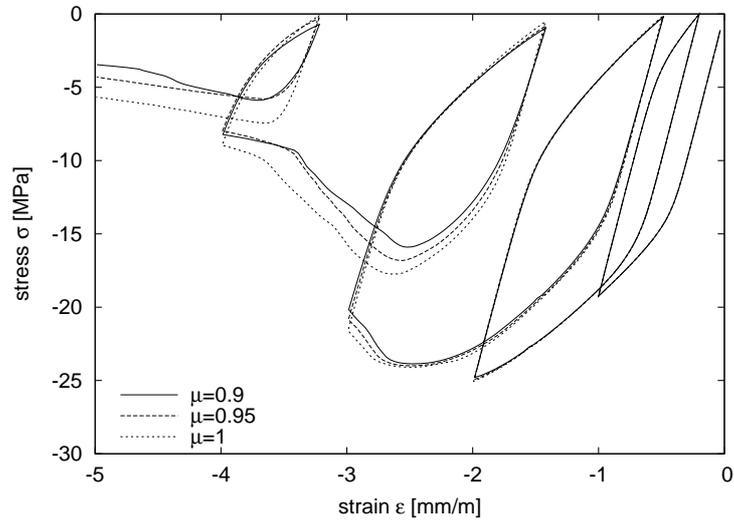,width=10cm}
  \end{center} 
  \caption{Stress-strain responses for cyclic loading for $\mu=1$, $0.95$ and $0.9$ of the mortar phase.}
  \label{fig:ldHpMortarCyclig}
\end{figure}

Furthermore, the influence of the permanent displacements in the mortar phase is studied. Again, the same values of $\mu = 1$, $0.95$, $0.9$ were used. The average stress-strain relations for monotonic and cyclic loading are presented in Figures~\ref{fig:ldHpMortarMono}~and~\ref{fig:ldHpMortarCyclig}.
The ratio of the mortar phase has almost no influence on the pre-peak regime, since almost all nonlinearities result from the nonlinearities at the ITZs. However, in the post-peak regime nonlinear displacements occur in the mortar phase.

\begin{figure}
  \begin{center}
    \epsfig{file=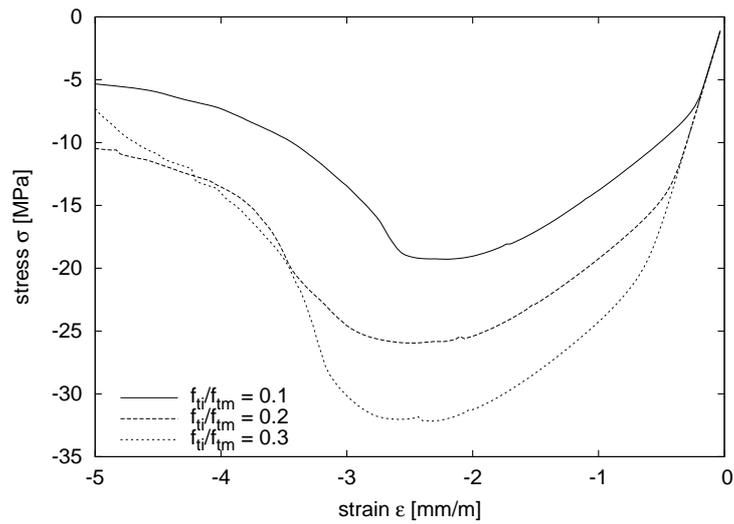,width=10cm}
  \end{center}
  \caption{Stress-strain responses for monotonic loading for interface-mortar strength ratios of $f_{\rm ti}/f_{\rm tm} = 0.1$, $0.2$ and $0.3$.}
  \label{fig:ldIntStrengthMono}
\end{figure}

\begin{figure}
  \begin{center}
    \epsfig{file=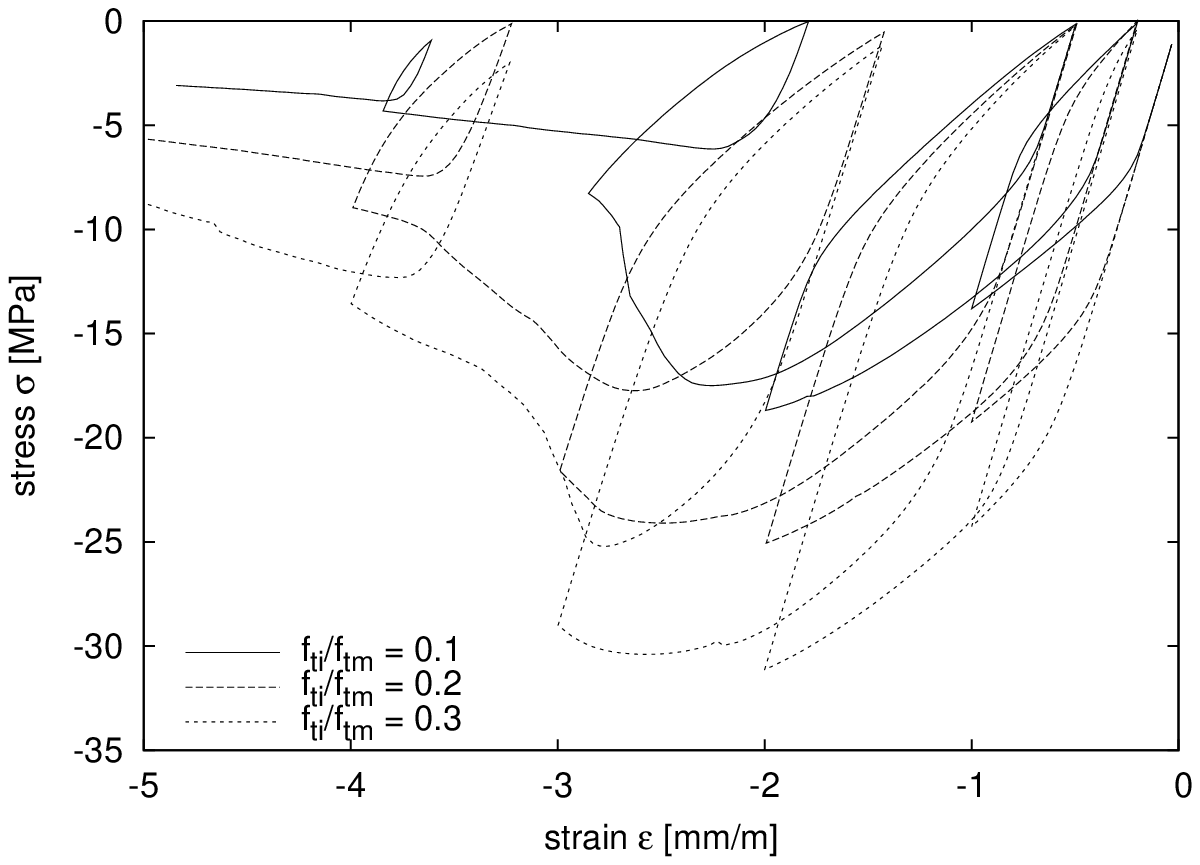,width=10cm}
  \end{center}
  \caption{Stress-strain responses for cyclic loading for interface-mortar strength ratios of $f_{\rm ti}/f_{\rm tm} = 0.1$, $0.2$ and $0.3$.}
  \label{fig:ldIntStrengthCyclic}
\end{figure}

Finally, the influence of the ratio of the strength of ITZ and mortar was investigated. Three strength ratios of  $f_{\rm{i}}/f_{\rm{m}}=0.1$, $f_{\rm{i}}/f_{\rm{m}}=0.2$ and $f_{\rm{i}}/f_{\rm{m}}=0.3$ were considered. The average stress-strain relations for monotonic and cyclic loading are shown in Figures~\ref{fig:ldIntStrengthMono}~and~\ref{fig:ldIntStrengthCyclic}. 

The strength ratio of ITZ and mortar has a strong influence on the monotonic and cyclic response. 
The compressive strength increases with an increased ITZ/mortar strength ratio, since the strength of the weak zones (ITZs) is increased. 
For the cyclic response, the size of the hysteresis loops decreases with increasing interface strength, since less permanent displacements take place at the interfaces between mortar and aggregates.

\section{Conclusions}
In the present work a meso-scale modelling approach developed in \cite{GraRem08} was used for a parametric study of the modelling of concrete subjected to cyclic compression.
Concrete was idealised by a three-phase composite consisting of mortar, aggregates and interfacial transition zones.
Five parameters were investigated: aggregate volume fraction, aggregate size, ratio of the permanent and total inelastic displacements in the ITZ and mortar phase ,and the ratio of the strength of the interfacial transition zone and the mortar phase.
The influences of the different parameters are summarised as following:

\begin{itemize}
\item{A decrease of the volume fraction of aggregate increases the ultimate strength of concrete, since fewer aggregates correspond to a reduction of ITZs, which weaken the material.
A decrease of volume fraction reduces also the size of the hysteresis loops, since the number of ITZs, at which permanent displacements take place, are reduced.
Localised permanent displacements are the main reason for the occurrence of hysteresis loops in the present meso-scale modelling approach.}

\item{The size range of aggregates does not strongly influence the response of concrete subjected to monotonic and cyclic loading.}

\item{A decrease of the amount of permanent displacements in the ITZ phase results in a reduction of the size of the hysteresis loops.}

\item{A decrease of the amount of permanent displacements in the mortar phase results in a decrease of the size of the loops in the post-peak regime, in which localised permanent displacements occur in the mortar phase.}

\item{Finally, the ratio of the strength of the interfacial transition zone and the mortar has a strong influence on both the monotonic and cyclic response.
An increase of the strength of ITZ leads to an increase of the compressive strength, since the amount of permanent displacements in the ITZs is reduced.
Furthermore, the size of the hysteresis loops is decreased as well.}
\end{itemize}

\section*{Acknowledgements}
The simulations were performed with the object-oriented finite
element package OOFEM \cite{Pat99,PatBit01} extended by the
present authors. The mesh has been prepared with the mesh generator
Triangle~\cite{shew96}.
\bibliographystyle{unsrt}
\bibliography{general}

\end{document}